\documentclass[sigconf]{acmart}

\usepackage{colortbl}      
\usepackage[table]{xcolor}
\usepackage{amsmath}
\usepackage{latexsym}
\usepackage{multirow}
\usepackage{enumitem}
\usepackage{makecell}
\usepackage{mdframed} 
\usepackage{ragged2e} 
\usepackage{algorithm}
\usepackage{algpseudocode}
\usepackage{arydshln}
\usepackage{enumitem}

\newcommand{\shrink}{\vspace*{-.9\baselineskip}}

\newcommand{\proposedmethod}{LegacyTranslate}
\newcommand{\AgentOne}{Initial Translation Agent}
\newcommand{\AgentTwo}{API Grounding Agent}
\newcommand{\AgentThree}{Refinement Agent}

\AtBeginDocument{%
  }

\setcopyright{acmlicensed}
\copyrightyear{2018}
\acmYear{2018}
\acmDOI{XXXXXXX.XXXXXXX}
\acmConference[Conference acronym 'XX]{Make sure to enter the correct
  conference title from your rights confirmation email}{June 03--05,
  2018}{Woodstock, NY}
\acmISBN{978-1-4503-XXXX-X/2018/06}

\begin{document}

\title{LegacyTranslate: LLM-based Multi-Agent Method for Legacy Code Translation}

\author{Zahra Moti}

\affiliation{%
  \institution{Radboud University}
  \city{Nijmegen}
  \country{The Netherlands}}
\email{zahra.moti@ru.nl}

\author{Heydar Soudani}
\affiliation{%
  \institution{Radboud University}
  \city{Nijmegen}
  \country{The Netherlands}}
\email{heydar.soudani@ru.nl}

\author{Jonck van der Kogel}
\affiliation{%
  \institution{ING}
  \city{Amsterdam}
  \country{The Netherlands}}
\email{jonck.van.der.kogel@ing.com}

\renewcommand{\shortauthors}{Moti et al.}

\begin{abstract}
Modernizing large legacy systems remains a major challenge in enterprise environments, particularly when migration must preserve domain-specific logic while conforming to internal architectural frameworks and shared APIs. Direct application of Large Language Models (LLMs) for code translation often produces syntactically valid outputs that fail to compile or integrate within existing production frameworks, limiting their practical adoption in real-world modernization efforts.
In this paper, we propose \proposedmethod, a multi-agent framework for API-aware code translation, developed and evaluated in the context of an ongoing modernization effort at a financial institution migrating approximately 2.5 million lines of PL/SQL to Java.
The core idea is to use specialized LLM-based agents, each addressing a different aspect of the translation challenge. Specifically, \proposedmethod\ consists of three agents:
\AgentOne\ produces an initial Java translation using retrieved in-context examples; 
\AgentTwo\ aligns the code with existing APIs by retrieving relevant entries from an API knowledge base; 
and \AgentThree\ iteratively refines the output using compiler feedback and API suggestions to improve correctness.
Our experiments show that each agent contributes to better translation quality. The \AgentOne\ alone achieves 45.6\% compilable outputs and 30.9\% test-pass rate. 
With \AgentTwo\ and \AgentThree, compilation improves by an additional 8\% and test-pass accuracy increases by 3\%.
These results highlight that without explicit API grounding, generated code rarely compiles in production frameworks; that retrieval quality directly affects functional correctness; and that model capacity plays a decisive role in handling framework-level constraints. These observations provide actionable guidance for teams considering LLM-assisted legacy migration.

\end{abstract}




\maketitle

\section{Introduction}
Modernizing large legacy systems remains a pressing concern for many enterprises that continue to depend on applications written decades ago in languages such as PL/SQL or COBOL~\cite{Hans25Automated}. 
As expertise in these technologies declines and architectural requirements evolve, organizations are increasingly seeking to migrate legacy codebases to modern platforms such as Java~\cite{Solovyeva25Leveraging, lachaux2020unsupervised}. 
The system studied in this paper is a large-scale legacy application at a European financial institution, comprising approximately 2.5 million lines of PL/SQL code. It lacks consistent documentation and is currently being migrated to a modern Java-based architecture that relies on shared libraries and internal APIs, with which all translated code must comply. Although recent advances in automated and LLM-based code translation have shown promising results~\cite{yang2024exploring, lachaux2020unsupervised}, applying these techniques in real-world modernization settings remains challenging.
Most existing evaluations of code translation focus on widely used language pairs, such as Python and C++, where abundant training data and mature tooling are available~\cite{yuan2024transagent, zhu2022multilingual}. 
In contrast, enterprise modernization often involves niche legacy languages and domain-specific frameworks with limited public benchmarks~\cite{Solovyeva25Leveraging}. Moreover, prior work typically evaluates translation at the function level, whereas industrial migration requires transforming modules that interact with internal libraries, shared APIs, and architectural conventions~\cite{chang2023rtcoder, yang2024exploring}. These dependencies introduce constraints that are rarely captured in benchmark settings.

\begin{figure}[t]
    \centering
    \includegraphics[width=0.43\textwidth]{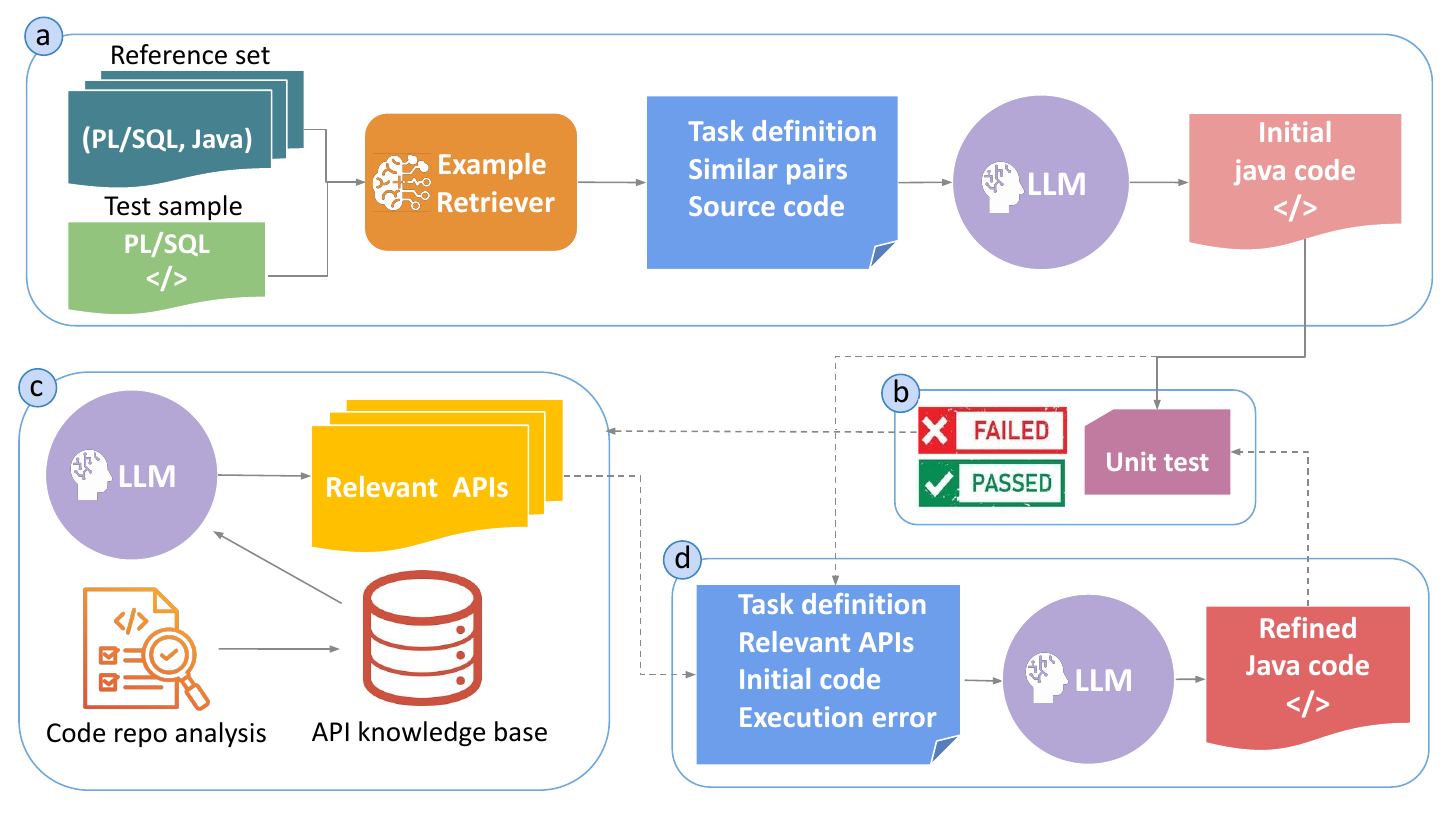}
    \shrink
    \caption{Overview of \proposedmethod\ for PL/SQL$\rightarrow$Java translation. 
(a) \AgentOne\ takes a test sample and reference set and generates an initial code using in-context examples retrieval.
(b) The code is evaluated using unit tests. 
(c) \AgentTwo\ shortlists relevant APIs from a knowledge base.
(d) \AgentThree\ iteratively corrects the code using API details and execution feedback. Solid lines = standard flow; dotted = executed only on failure.
}
\label{fig:overview}
\shrink
\end{figure}

In earlier work, Solovyeva et al.~\cite{Solovyeva25Leveraging} conducted a feasibility study investigating the use of LLMs for translating PL/SQL to Java within a similar modernization context. That study demonstrated that LLMs, guided by customized prompting strategies, can produce syntactically accurate translations with a degree of functional correctness. However, it was limited to 10 code pairs, did not account for the API and framework-level constraints of the target architecture, and had restricted access to unit tests for validation. These limitations left open whether LLM-based translation could produce outputs that compile and pass tests within the production framework at a larger scale.

In this paper, we present \proposedmethod, a multi-agent framework for API-aware legacy code translation that addresses these gaps. Rather than relying on a single LLM prompt, \proposedmethod\ decomposes the translation task into three coordinated stages:
1) \AgentOne, which performs the first-stage translation from the legacy language to the modern target language. This initial pass produces a draft version of the code that may still require refinement and API awareness, which are handled by the subsequent agents.
This agent is implemented using an in-context examples retriever that retrieves semantically similar code pairs to guide the LLM in the translation task.
2) \AgentTwo, which shortlists relevant APIs from the API knowledge base to ensure alignment with company-specific libraries.
3) \AgentThree, which iteratively corrects the code using compiler/test feedback and the selected APIs, producing a more compliant and executable translation.
Although we focus on translating PL/SQL to Java in this paper, the proposed approach is generalizable to other language-migration scenarios, which we leave for future work.
The full pipeline achieves 52.9\% compilation and 33.8\% test-pass rates. \AgentOne\ alone reaches 45.6\% compilation and 30.9\% test-pass, with \AgentTwo\ and \AgentThree\ contributing further gains. A central finding is that without explicit API grounding, virtually no generated code compiles within the target framework — highlighting the gap between syntactically valid LLM output and production-ready code. We also observe that translation quality varies substantially with both the LLM and retriever capacity.
The main contributions of this paper are:
\begin{enumerate}[leftmargin=*]
\item We propose \proposedmethod, a multi-agent code translation framework designed to translate legacy PL/SQL code into modern Java while incorporating company-specific APIs.
\item We demonstrate that each agent consistently contributes to improving translation quality, highlighting the effectiveness of multi-agent architectures in real-world code translation.
\item We evaluate the performance of \proposedmethod\ using various LLMs and retrievers, underscoring the importance of model selection for achieving high translation performance.
\end{enumerate}

\section{Related work}

\noindent\textbf{Code Translation.} 
Most prior work on code translation has focused on popular language pairs such as Java, Python, and C++, primarily targeting small, self-contained code units — typically functions or method-level snippets. Zhu et al.~\cite{zhu2022multilingual} proposed a multilingual pre-training approach that improves code translation by jointly training on code in multiple languages. Roziere et al.~\cite{lachaux2020unsupervised} introduced an unsupervised neural approach that learns to translate code between languages without parallel data using cross-lingual embeddings and denoising objectives. Szafraniec et al.~\cite{szafraniec2022code} proposed using compiler intermediate representations to guide code translation, enabling models to better preserve program semantics across languages. More recently, UniTrans~\cite{yang2024exploring} generates target-specific unit tests from source code and uses these tests to guide and verify the translation. However, such successes remain confined mainly to small code units and do not generalize to the class- or framework-level migrations faced in enterprise modernization.
Other studies confirm this gap. G-TransEval~\cite{Jiao23Taxonomy} categorizes translation tasks by complexity and finds that while models handle simple token- or syntax-level translations well, they struggle with more complex, framework- or algorithm-level tasks~\cite{xue2024escalating}. \citet{Deng25Enhancing} tackles project-specific code completion by retrieving from internal API knowledge, demonstrating the importance of API awareness, though their focus is on code completion task.

\noindent\textbf{Legacy Code Modernization in Industry.} While academic benchmarks focus on function-level translation between popular languages, industrial modernization involves migrating large codebases written in niche languages under real architectural constraints. Hans et al.~\cite{Hans25Automated} and Solovyeva et al.~\cite{Solovyeva25Leveraging} are among the few works addressing legacy translation in industrial settings, focusing on translating COBOL and PL/SQL to Java translation, respectively.

\noindent\textbf{Multi-Agent Collaboration.} Multi-agent systems enable groups of intelligent agents to coordinate and solve complex tasks at scale, shifting from isolated models toward collaboration-centric approaches~\cite{Tran25Collaboration}. Recently, TransAgent~\cite{yuan2024transagent} introduced multiple agent roles — including a code generator, syntax checker, and semantic analyzer — to iteratively refine translated code. However, it operates only on short snippets and does not address the framework-dependent needs of real modernization tasks, such as alignment with internal APIs and shared libraries. A clear gap remains in defining a multi-agent approach tailored for translating legacy code to modern code in an API-aware, enterprise setting.

\section{Methodology}

\setlength{\abovedisplayskip}{2pt} 
\setlength{\belowdisplayskip}{2pt} 

We propose an LLM-based multi-agent framework, \proposedmethod, to accelerate the migration of a legacy PL/SQL codebase into a modern Java-based architecture.  
The source code is a large, undocumented, and database-centric framework in which business logic is embedded directly within SQL code.  
As part of an ongoing modernization effort, engineers have already initiated the migration by designing a new architecture with well-defined interfaces and APIs that govern how new modules should be implemented.  
Building on this foundation, our approach leverages LLMs to translate PL/SQL code into Java while ensuring that the generated code conforms to the existing architectural and API constraints.
Figure~\ref{fig:overview} shows an overview of \proposedmethod.  
Starting from a PL/SQL code unit, the method produces a Java implementation that aligns with the existing architecture and APIs through three agents: \AgentOne, \AgentTwo, and \AgentThree. 
Each agent builds on the output and feedback of the previous one, progressively improving translation quality and compliance.

As shown in Figure~\ref{fig:overview} (a), the process begins with the \AgentOne.  
Given a PL/SQL test sample, it generates the initial Java code using retrieval-augmented prompting.  
An Example Retriever identifies semantically similar PL/SQL$\rightarrow$Java pairs from a reference set and incorporates them into the LLM prompt as few-shot demonstrations, alongside the task definition and the source code to be translated.  
The LLM then produces an initial Java draft, which is evaluated for \textit{compilation success} and \textit{test-case correctness} (Figure~\ref{fig:overview}b).
If the code fails to compile or pass tests, the process advances to the \AgentTwo\ (Figure~\ref{fig:overview}c).  
This agent identifies the APIs that are likely missing or misused in the generated code.  
To this end, the LLM is instructed to select relevant APIs from a curated knowledge base.  
The resulting shortlisted APIs are then passed to the \AgentThree\ to guide the next correction step.
Finally, the \AgentThree\ (Figure~\ref{fig:overview}d) iteratively improves the code through a correction loop.  
Using the relevant APIs from the previous step, together with compiler errors and unit-test feedback, the agent re-prompts the LLM to generate an improved code.  
This process repeats until the code compiles successfully, passes all associated test cases, or reaches a predefined termination criterion.  
We next describe each agent in detail.

\subsection{\AgentOne}

\proposedmethod\ begins with an initial translation generated by \AgentOne. The term ``initial'' emphasizes that this agent’s primary role is to introduce the translation task from the legacy language to the new one. 
\AgentOne\ is implemented using in-context examples retrieval. Specifically, we leverage a reference set of aligned PL/SQL$\rightarrow$Java pairs as exemplars and retrieve relevant examples to teach the agent how to perform the translation. Formally, given a reference set $D = \{(s_n, j_n)\}_{n=1}^N$ and a test PL/SQL query $s_t$, a retriever model $\mathcal{R}$ is used to obtain $D_k$, the top-$k$ semantically similar translation pairs from $D$:
$$D_k = \mathcal{R}(D, s_t),$$
where $k$ indicates the number of exemplars provided.
After retrieval, an LLM uses the retrieved exemplars to generate the Java translation. Formally, the LLM $\pi_{\theta}$ with parameters $\theta$ instructed by $D_k$ and $s_t$, the model produces the Java translation $j_t$:
$$j_t \gets \pi_{\theta}(D_k, s_t).$$

\renewcommand{\arraystretch}{1}
\begin{table}[t]
\centering
\setlength{\tabcolsep}{2.8pt}
\caption{Performance of API-aware code translation. 
The results indicate the effectiveness of each agent. 
}
\label{tab:main}
\shrink
\begin{tabular}{l|cccc}
\hline
\textbf{Method} &  
\makecell{\textbf{Structural} \\ \textbf{Validity (\%)}} &
\makecell{\textbf{Compilation } \\ \textbf{Rate (\%)}} &
\makecell{\textbf{Test Pass} \\ \textbf{Rate (\%)}}  \\
\hline\hline
LLM-only                      & 42.6 & 0.0  & 0.0  \\  
LLM+Rand. Examples            & 98.5  & 1.5  & 0.0  \\ 
\textbf{\proposedmethod}      & \textbf{100}  & \textbf{52.9} & \textbf{33.8} \\ 
- \textit{w/o API \& Refine}  & 98.5  & 45.6 & 30.9 \\ 
\hline
\end{tabular}
\shrink 
\end{table}

\subsection{\AgentTwo}~\label{sec:api-agent}
Cases where the initial translation fails to compile can be improved by leveraging compiler feedback and API-level adaptation.
\AgentTwo\ is responsible for providing relevant APIs. We construct a curated \textit{API knowledge base} from the project’s architected Java libraries—specifically the shared APIs and interfaces that define how new Java components should be implemented in the modernized framework. Given such a repository, we run a static analysis that traverses Java source files and extracts each API’s key attributes: its class, signature (name, parameters, return type), body, and file location. We then generate brief descriptions for these entries using an LLM.
Using this knowledge base—along with the initial generated code and the test feedback—\AgentTwo\ prompts an LLM to shortlist the APIs most likely needed to fix the errors in the initial code. Formally, given the API knowledge base $A$, the Java translation $j_t$, and the compiler error messages $E_t$, the LLM $\pi_{\theta}$ is prompted to produce a shortlist of relevant APIs $A_t$:
$$A_t \gets \pi_{\theta}(A, E_t, j_t).$$
The length of the shortlist is determined by the LLM $\pi_{\theta}$, based on the content of $j_t$.

\subsection{\AgentThree}
Compiler errors are effective signals for improving code translations~\cite{Zhangqian24Iterative}.
Thus, \AgentThree\ is designed to perform iterative code refinement using compiler error messages together with the shortlisted APIs.
Formally, at each iteration, the agent compiles the currently generated code ${j_t}^m$ and collects the resulting error messages ${E_t}^m$.
The LLM $\pi_{\theta}$ is then prompted with 
${j_t}^m$, the shortlisted APIs $A_t$, and the error feedback ${E_t}^m$ to generate an improved translation ${j_t}^{m+1}$:
$$j_{t}^{m+1} \gets \pi_{\theta}\!\left(A_t, {E_t}^m, {j_t}^m\right); \quad m = 0, \dots, M_t.$$
Here, ${j_t}^0$ is the output of \AgentOne, $j_t$, and ${E_t}^0$ is ${E_t}$.
The refinement loop continues until the code compiles successfully or until no additional test cases pass after a refinement step.
Thus, the value of $M_t$ is not predefined and varies for each test sample.
\section{Experimental Setup}

\noindent
\textbf{\textit{Models}}:
As the backbone LLM, we use \textit{Qwen2.5-Coder-14B}~\cite{hui2024qwen2}, an instruction-tuned model optimized for code-related tasks.
For retrieval, we employ \textit{SFR-Mistral}~\cite{SFRAIResearch2024}, a dense retriever with 4096 embedding vector size, following prior work on retrieval-augmented code generation~\cite{wang2024coderagbench}.
We set the number of retrieved exemplars to $k=3$, which provides the best trade-off between relevance and prompt length.
In \AgentTwo, for generating descriptions during the construction of the API Knowledge Base, we use \textit{Claude 3.7 Sonnet}.
All experiments are conducted on a single NVIDIA A100 GPU with 40 GB of memory, requiring approximately $\sim${5} GPU hours in total.

\vspace{0.5em}
\noindent
\textbf{\textit{Evaluation Metrics}}:
We evaluate the generated Java code using three standard metrics:
\textit{Structural Validity (SV)}, \textit{Compilation Rate (CR)}, 
and \textit{Test Pass Rate (TPR)}. 
SV measures whether the output forms a valid Abstract Syntax Tree. 
CR indicates whether the code compiles when executed in our unit-test environment. 
TPR reflects functional correctness, measured as the proportion of tests passed by each compiled output.
To assess the example retrieval quality, we report 
\textit{NDCG@3}, \textit{MRR@3} (Mean Reciprocal Rank), and \textit{Recall@3}.  
NDCG@3 measures overall ranking quality,  
MRR@3 evaluates how early the first relevant example appears,  
and Recall@3 indicates whether any relevant example is retrieved within the top~3.

\vspace{0.5em}
\noindent
\textbf{\textit{Baselines}}:
We compare \proposedmethod\ with two baselines. In \textit{LLM-only}, the model directly translates the test sample without example pairs or API information. In \textit{LLM+Random Examples}, three randomly selected PL/SQL-to-Java pairs are included as in-context examples, matching the number of retrieved examples used by \proposedmethod.
We do not include the prompting strategy of Solovyeva et al.~\cite{Solovyeva25Leveraging} as a baseline, as their prompts are not publicly available.

\vspace{0.5em}
\noindent
\textbf{\textit{Dataset}}:
We evaluate our method on an internally curated dataset of 68 aligned PL/SQL$\rightarrow$Java pairs, representative of the module types found across the legacy codebase. Each PL/SQL snippet was manually migrated by engineers, providing a ground-truth Java reference. The PL/SQL code ranges from 65 to 292 lines (mean 101.7 lines of code). Each sample includes multiple unit tests written by engineers and executed within the actual production framework. These tests validate the generated code at multiple levels, including structural validity, successful compilation within the framework, API conformance, and functional correctness.
We also maintain a separate reference set of 68 aligned pairs used by the example retriever to provide in-context examples, as well as an API knowledge base containing 80 entries extracted from the shared Java libraries of the target architecture (see Section~\ref{sec:api-agent}). To the best of our knowledge, no public benchmark currently supports realistic PL/SQL$\rightarrow$Java translation or API-dependent migration. Our evaluation relies on real production code from an active modernization project, providing a realistic assessment.

\vspace{0.5em}
\noindent
\textbf{\textit{Prompts}}: We follow \citet{Solovyeva25Leveraging} to design the input prompt. Specifically, the prompt consists of three key components: (1) a description of the target Java architecture, including its shared interfaces; (2) retrieved examples of PL/SQL–to–Java translation pairs; and (3) the PL/SQL code snippet serves as the input for translation.

\section{Results}

\begin{figure}[t]
  \centering
  \includegraphics[width=0.46\textwidth]{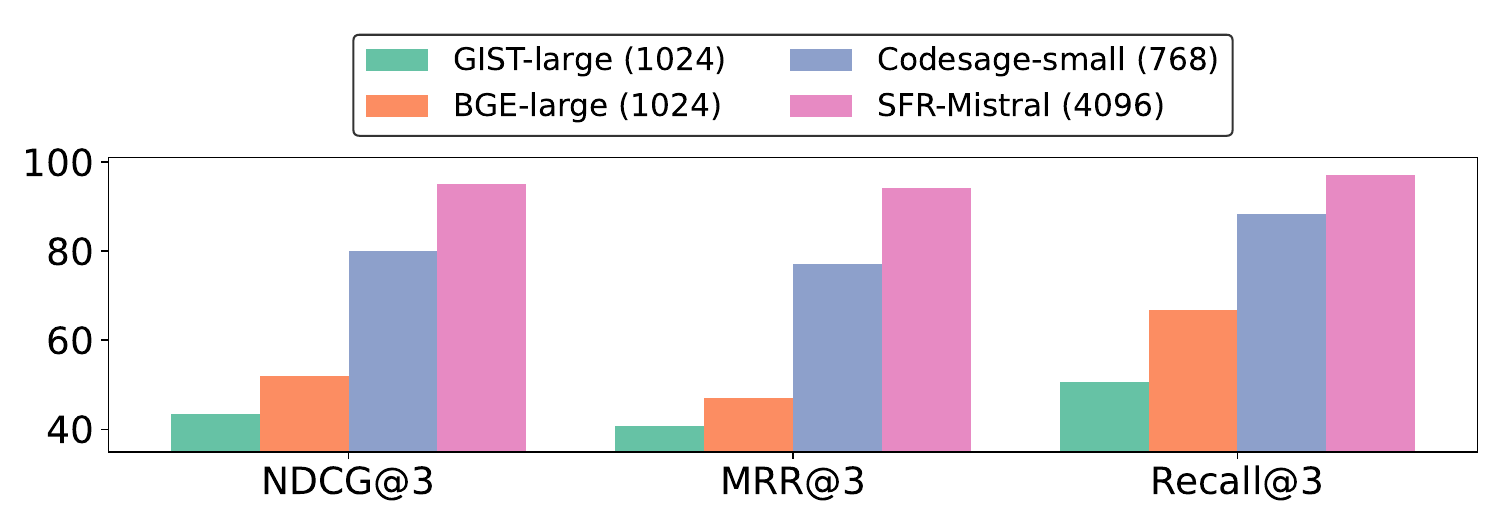}
  \shrink
  \caption{Performance of different retrieval models on retrieving examples in \AgentOne.}
  \label{fig:retrievers}
  \shrink
\end{figure}

\textbf{\textit{RQ1}: How does \proposedmethod\ perform on the API-aware PL/SQL $\rightarrow$Java translation?}
Table \ref{tab:main} reports the performance of \proposedmethod\ compared with other baselines. When directly prompting the LLM to perform translation (\textit{LLM-only} baseline), none of the samples compile or pass the test cases. This is because the target framework requires code to extend specific base classes and use internal APIs; without any examples or architectural guidance, the LLM generates generic Java that is syntactically valid but fails to compile within the production environment.
Adding a few random examples to the input prompt (LLM+Random Examples baseline) substantially increases SV from 42.6\% to 98.5\%, indicating that even unrelated examples help the LLM produce more structurally consistent outputs. However, the CR only increases marginally, from 0.0\% to 1.5\%, showing that random examples do not help the model to learn the patterns.
In contrast, \proposedmethod\ improves API-aware code translation by a large margin. The SV reaches 100\%, and, more importantly, the CR and TPR rise to 52.9\% and 33.8\%, respectively, showing the effectiveness of our multi-agent system. We also evaluate the output of \AgentOne\ alone reported in the “\textit{w/o API \& Refine}” row. This shows the contribution of each agent individually: using only \AgentOne\ already improves CR and TPR to 45.6\% and 30.9\%, and adding \AgentTwo\ and \AgentThree\ leads to further gains. 
A second refinement round yielded an additional 3\% improvement before convergence, suggesting that further gains may require stronger models or deeper architectural knowledge.  
Overall, these results show that our multi-agent approach effectively enables API-aware translation from a legacy language to a modern one, demonstrating its practical feasibility.

\renewcommand{\arraystretch}{1}
\begin{table}[t]
\centering
\small
\setlength{\tabcolsep}{2.6pt}
\caption{Performance of \AgentOne\ with different LLMs. 
}
\label{tab:llm_comparison}
\shrink
\begin{tabular}{l|cccc}
\hline
\textbf{Method} & 
\makecell{\textbf{Structural} \\ \textbf{Validity (\%)}} &
\makecell{\textbf{Compilation } \\ \textbf{Rate (\%)}} &
\makecell{\textbf{Test Pass} \\ \textbf{Rate (\%)}} &
\makecell{\textbf{Generation} \\ \textbf{Time (s)}} \\

\hline\hline
Qwen2.5-14B     & \textbf{98.6}   & \textbf{45.58} & \textbf{30.88} & 32.26 \\
CodeLlama-13B  & 8.83 & 0.0   & 0.0   & 40.26 \\
Qwen2.5-7B     & 7.36 & 0.0   & 0.0   & 19.27 \\
CodeGemma-7B   & 0.0   & 0.0   & 0.0   & \textbf{6.50}  \\
\hline
\end{tabular}
\shrink
\end{table}

\vspace{0.5em}
\noindent
\textbf{\textit{RQ2}: How does the choice of retriever and LLM affect \proposedmethod\ performance?}
To assess the impact of retriever and LLM selection, we focus on \AgentOne, which includes both retrieval and LLM models. Figure~\ref{fig:retrievers} presents the example-retrieval performance of different retrievers. The embedding dimension is indicated next to each model name; among them, \textit{SFR-Mistral} has the largest embedding size (4096). It achieves the highest retrieval performance, substantially outperforming all other candidates. This finding aligns with prior work~\cite{wang2024coderagbench}, which showed that high-dimensional dense retrievers provide more semantically aligned examples for code-related tasks.
After fixing \textit{SFR-Mistral} as the retriever, we evaluated several LLMs for code translation, with results reported in Table~\ref{tab:llm_comparison}. \textit{Qwen2.5-14B} delivers the best performance, consistently producing compilable, test-passing code with moderate latency (~32s per sample).  
Smaller models such as \textit{Qwen2.5-7B}, \textit{CodeLlama-13B}, and \textit{CodeGemma-7B} frequently fail to generate valid or compilable outputs. These models lack sufficient capacity to handle the long, context-heavy prompts required by the task, resulting in structurally invalid or non-compilable outputs.
Overall, translation quality scales with model capacity: larger, code-specialized LLMs yield higher correctness and better framework alignment, though at the cost of longer generation time.

\section{Conclusions and Future Work}
This paper introduces a novel code translation method for API-aware translation from the legacy language PL/SQL to the modern language Java—a highly practical and in-demand task in industry. \proposedmethod\ is a multi-agent framework consisting of three agents:  
\AgentOne\ generates the initial translation, \AgentTwo\ aligns it with existing APIs, and \AgentThree\ refines it using compiler feedback.
Our results show that \proposedmethod\ is effective, generating 52.9\% compilable code and demonstrating the feasibility of modernizing legacy systems.  
However, API-aware legacy translation remains limited by the lack of public datasets or fine-tuned models, which we encourage future work to tackle.  
Moreover, our multi-agent framework can be extended to other languages and enriched with additional agents.


\bibliographystyle{ACM-Reference-Format}
\balance
\bibliography{main}

\newpage
\appendix

\end{document}